\documentclass[copyright]{eptcs}
\providecommand{\event}{J.~Dassow, G.~Pighizzini, B.~Truthe (Eds.): 11th International Workshop\\
on Descriptional Complexity of Formal Systems (DCFS 2009)} 
\providecommand{\volume}{3}
\providecommand{\anno}{2009}
\providecommand{\firstpage}{3} 
\providecommand{\eid}{1}
\usepackage{breakurl}        

\input{dcfs.tex}
\usepackage{gastex}
\usepackage{extarrows}

\def\word{{\sigma}}

\def\M{{\mathcal M}}
\def\N{{\mathcal N}}
\def\S{{S}}

\def\Pr{{\sf Pr}}

\def\Q{{Q}}
\def\Acc{{\sf Acc}}

\def\inf{{\sf inf}}

\def\L{{\mathcal L}}

\def\P{{\mathcal P}}

\def\eqdef{\stackrel{\text{\rm def}}{=}}

\def\T{{\mathcal T}}

\def\<{\langle}
\def\>{\rangle}

\newcommand{\overto}[1]{\xlongrightarrow{#1}}

\title{Probabilistic Automata over Infinite Words: \\
  Expressiveness, Efficiency, and Decidability\thanks{The collaboration of the authors is supported
  by a bilateral PROCOPE-project.}}
\author{Christel Baier
\institute{Technische Universit\"at Dresden --
Fakult\"at f\"ur Informatik\\
  Germany}
\email{baier@tcs.inf.tu-dresden.de}
\and
Nathalie Bertrand
\institute{INRIA Rennes Bretagne Atlantique\\France}
\email{nathalie.bertrand@irisa.fr}
\and
Marcus Gr\"o{\ss}er
\institute{Technische Universit\"at Dresden -- 
Fakult\"at f\"ur Informatik\\
  Germany}
\email{groesser@tcs.inf.tu-dresden.de}
}
\def\titlerunning{Probabilistic Automata over Infinite Words}
\def\authorrunning{C.~Baier, N.~Bertrand, M.~Gr\"o{\ss}er}
\begin{document}
\maketitle

\begin{abstract}
Probabilistic $\omega$-automata are
variants of nondeterministic automata for infinite
words where all choi\-ces are resolved by probabilistic
distributions. Acceptance of an infinite input word
can be defined in different ways:  by requiring that 
(i) the probability for the 
accepting runs is positive (probable semantics), 
or
(ii) almost all runs are accepting (almost-sure semantics), 
or 
(iii) the probability measure of the accepting runs
 is greater than a certain threshold (threshold semantics).
The underlying notion of an accepting run can be defined
as for standard $\omega$-automata by means of a
B\"uchi condition or other acceptance conditions, e.\,g.,
Rabin or Streett conditions.
In this paper, we put the main focus on the probable semantics and
provide a summary of the fundamental
properties of probabilistic $\omega$-automata 
concerning expressiveness, efficiency,  and 
decision problems. 
\end{abstract}
\section{Introduction}

While classical finite automata can serve to recognize languages
over finite discrete structures, 
$\omega$-au\-to\-ma\-ta are acceptors 
for languages consisting of infinite objects.
They have been applied in various research areas, including
the verification of reactive systems and reasoning about
infinite games and decision problems for certain logics.
Many variants of $\omega$-automata have been studied 
in the literature
that can be classified according to 
their inputs (e.\,g., words or trees),
their acceptance conditions
(e.\,g., B\"uchi, Rabin, Streett, Muller or parity acceptance)
and their branching structure 
(e.\,g., deterministic, nondeterministic, or alternating).
We refer to \cite{Thomas97handbook,GraedelThomasWilke02}
for an overview of automata over infinite objects.

Probabilistic variants of $\omega$-automata
for languages over infinite words have been recently  
introduced. Their syntax is roughly the 
same as for probabilistic finite automata
(PFA) \cite{Rabin63,paz71}, i.\,e.,
they are finite-state automaton where for each state $q$ 
and input letter a probability distribution specifies the probabilities
for the sucessor states. Furthermore, they are equipped with an acceptance
condition as in nondeterministic $\omega$-automata.
The accepted language of a probabilistic $\omega$-automata can be defined
by imposing a condition on the acceptance probability for the input words.
Under the \emph{probable semantics},
acceptance of an infinite word $\sigma \, = \, a_1 \, a_2 \, a_3 \ldots$
requires that the generated sample run for $\sigma$
(i.\,e., sequence of states that are passed in the automaton
while reading $\sigma$ letter by letter)
meets the acceptance condition with positive probability.
The probable semantics is in the spirit of  nondeterministic automata where the accepted words are those
words that have at least one accepting run.
The \emph{almost-sure semantics} of a probabilistic $\omega$-automata
can be understood as the probabilistic counterpart to universal
automata as it requires that the accepting runs have probability 
measure~1, i.\,e., almost all runs are accepting.
The \emph{threshold semantics} follows the concept of PFA
and deals with a fixed threshold $\lambda \in ]0,1[$ and
classifies an input word $\sigma$ to be accepted if the 
probability of the
accepting runs for $\sigma$ is greater than $\lambda$.

The different semantics yield different classes
of recognizable languages over infinite words.
Most powerful is the threshold semantics which covers
the class of $\omega$-regular languages, but also non-$\omega$-regular
languages. Given the results for PFA which are known
to be more expressive than standard finite automata, this is not
surprising. 
While PFA with the probable semantics agree with
ordinary nondeterministic automata, 
probabilistic automata with B\"uchi or other standard 
acceptance conditions and the probable semantics are
strictly more expressive than their nondeterministic counterparts.
Furthermore, there are languages $L_n$ that are recognizable by
probabilistic B\"uchi automata of linear size, while smallest
nondeterministic $\omega$-automata for $L_n$ 
have exponentially many states.
For nondeterministic $\omega$-automata it is well-known
that B\"uchi acceptance is as powerful as, e.\,g.,
Streett or Rabin acceptance, but the transformations
from nondeterministic Streett automata to nondeterministic B\"uchi
automata can cause an exponential blow-up
\cite{SafraVardi89}. In contrast, 
there is a polynomial transformation from
probabilistic B\"uchi to probabilistic Streett automata,
both under the probable semantics.
Concerning the standard 
composition operators (union, intersection and complementation), 
the class of languages that are recognizable by 
probabilistic $\omega$-automata under the probable semantics enjoys the same properties
as the class of $\omega$-regular languages. 
Both are closed under all three operators. 
Union and intersection
can easily be realized by means of sum and product constructions, 
respectively. Complementation, however, is ``difficult'' and
relies on a complex powerset construction that  
can cause an exponential blow-up.
The price we have to pay for the extra power of
probabilistic $\omega$-automata under the probable semantics is that
all relevant decision problems (checking emptiness, universality 
or equivalence) are undecidable.
The undecidability results for PBA have several important consequences.
First,  the concept of PBA is
not adequate for solving algorithmic problems that are related to the
emptiness or universality problems. This, e.\,g., applies to 
the verification of
nondeterministic systems against PBA-specifications.
Second, PBA can be viewed as a special instance of partially-observable
Markov decision processes (POMDPs) which are widely used in various areas,
including robotics and stochastic planning (see, e.\,g.,
\cite{Sondik71thesis,PapTsi87,Lovejoy91})
and the negative results established for PBA yield the
undecidability of various verification problems for POMDPs.

For probabilistic B\"uchi automata with the almost-sure semantics
we obtain a completely different picture.
They are less powerful and even do not cover the full class of
$\omega$-regular languages, but still can accept languages that
are not $\omega$-regular. However, the emptiness and universality
problem are decidable for them. Furthermore, the class of languages 
that can be accepted by an almost-sure PBA is closed under union 
and intersection, but not under 
complementation. \vspace{-2mm}

\paragraph*{\it Organization.}
In Section \ref{sec:prob-omega-aut}, we briefly 
recall the definition of
nondeterministic $\omega$-automata with B\"uchi, Rabin or Streett acceptance
conditions and introduces their
probabilistic variants and the probable, almost-sure and threshold 
semantics.
The following three sections mainly deal with
probabilistic automata under the probable semantics.
Results on the expressiveness and efficiency of probabilistic
B\"uchi, Rabin and Streett automata are summarized in 
Section \ref{section:express}.
Composition operators for PBA under the probable semantics
are addressed
in Section~\ref{section:comp}.
Decision problems
for PBA and the relation to POMDPs
will be discussed in Section~\ref{section:dec}.
Section~\ref{section:almost}
summarizes the main results for the almost-sure
and threshold semantics. 
Finally, 
Section~\ref{section:conc} contains some concluding remarks.

The material of this paper is a summary of the results presented
in the papers~\cite{BG-lics05,BaiBerGroe08}. Further details
can be found there and in the thesis by Marcus Gr\"o{\ss}er~\cite{GroesserThesis}.


\section[Nondeterministic and probabilistic omega-automata]{Nondeterministic and probabilistic $\omega$-automata}
\label{sec:prob-omega-aut}

Throughout the paper, we assume some familarity with 
classical nondeterministic
automata over finite or infinite words and refer 
to \cite{Thomas97handbook,GraedelThomasWilke02} for details. 
We first recall some
basic concepts of
nondeterministic $\omega$-automata and then
 adapt these concepts to the probabilistic setting.

\begin{definition}[Nondeterministic $\omega$-automata] 
\label{def:nondet-omega-automata}\ \\
A nondeterministic $\omega$-automaton is a tuple
$\N = (\Q, \Sigma, \delta, \Q_0, \Acc), 
$
where
 \begin{itemize}
 \item  $\Q$ is a finite nonempty set of states,
 \item  $\Sigma$ is a finite nonempty input alphabet,
 \item $\delta : \Q \times \Sigma \, \to \, 2^\Q$
   is a \emph{transition function}, 
 \item $\Q_0 \subseteq \Q$ 
   is the set of {\em initial states},
 \item  $\Acc$ is an {\em acceptance condition}
   (which will be explained below).
 \end{itemize}
$\N$ is called deterministic if
$|\Q_0|=1$ and $|\delta(q,a)|=1$ for all $q\in Q$ and $a\in \Sigma$.
 \end{definition}

The intuitive operational behavior of a nondeterministic $\omega$-automaton
$\N$ for an infinite
input word $\sigma \ = \ a_1 \, a_2 \, a_3\ldots\in \Sigma^\omega$
is as follows. The computation starts in a 
nondeterministically chosen initial state
$q_0\in \Q_0$. Then, $\N$ attempts to read the first letter $a_1$
in state $q_0$. If $q_0$ does not have an outgoing $a_1$-transition
(i.\,e., $\delta(q_0,a_1)=\emptyset$) then the automaton rejects.
Otherwise, the automaton reads the first letter~$a_1$ and chooses
nondeterministically
 some state $q_1\in \delta(q_0,a_1)$. 
It then attempts to read the remaining word
$a_2 \, a_3 \ldots$ from state $q_1$. That is, the automaton rejects if
$\delta(q_1,a_2)=\emptyset$. Otherwise the automaton reads letter
$a_2$ and moves to some state $q_2\in \delta(q_1,a_2)$, and so on.
Any maximal state-sequence $\pi \ = \ 
  q_0 \, q_1 \, q_2 \, \ldots$ that can be
obtained in this way is called a {\em run} for $\sigma$.
We write $\inf(\pi)$ to denote the set of states $p\in \Q$ 
that appear infinitely often in $\pi$.
Each finite run $q_0\, q_1 \ldots q_i$ 
(where $\N$ fails to read letter $a_{i+1}$ in the last 
state $q_{i}$ because $\delta(q_i,a_{i+1})$ is empty)
is said to be \emph{rejecting}.
The acceptance condition $\Acc$ 
imposes a condition on infinite runs and declares which of the infinite
runs are {\em accepting}. Several acceptance conditions are known
for nondeterministic $\omega$-automata.
We will consider three types of acceptance conditions:
\begin{description}
\item [{\bf B\"uchi:}]
   A B\"uchi acceptance condition $\Acc$ is a subset $F$ of $\Q$.
  The elements in $F$ are called {\em final} or {\em accepting states}.  
   An infinite run $\pi\ = \ q_0 \, q_1\, q_2\ldots$ is called
  (B\"uchi) accepting if $\pi$ visits $F$ infinitely often, i.\,e.,
   $\inf(\pi) \cap F\not= \emptyset$.
\item [{\bf Streett:}]
  A Streett acceptance condition $\Acc$ is a finite set of pairs
  $(H_l,K_l)$ consisting of subsets $H_l,K_l$ of $\Q$, i.\,e.,
  $\Acc = \{(H_1,K_1),\ldots,(H_\ell,K_\ell)\}$.
  An infinite run
   $\pi \ = \ q_0 \, q_1\, q_2\ldots$ is called
  (Streett) accepting if for each $l\in \{1,\ldots,\ell\}$
  we have:
   $\inf(\pi)\cap H_l \neq \emptyset$ or
   $\inf(\pi) \cap K_l = \emptyset$.
\item [{\bf Rabin:}]
  A Rabin acceptance condition $\Acc$ is syntactically the same as
  a Streett acceptance condition, i.\,e., a finite set 
  $\Acc = \{(H_1,K_1),\ldots,(H_\ell,K_\ell)\}$ where
   $H_l,K_l \subseteq \Q$ for $1 \leq l \leq \ell$.
  An infinite run
   $\pi \ = \ q_0 \, q_1\, q_2\ldots$ is called
  (Rabin) accepting if there is some $l\in \{1,\ldots,\ell\}$
  such that 
   $\inf(\pi) \cap H_l = \emptyset$ and
   $\inf(\pi) \cap K_l \neq \emptyset$.
\end{description}
Using LTL-like notations,
a Streett condition can be understood as a
strong fairness condition and a Rabin condition as its dual. 
\begin{align*}
& \bigwedge\limits_{1 \leq l\le \ell} 
   (\Box \Diamond K_l \to \Box \Diamond H_l) \quad\mbox{(Streett)} \\
& \bigvee\limits_{1 \leq l\le \ell}
   (\Box \Diamond K_l \wedge \Diamond \Box \neg H_l) \quad \mbox{(Rabin)}
\end{align*}
Clearly, a B\"uchi acceptance
condition $F$
can be viewed as a special case of a Streett and Rabin condition
with a single acceptance pair, namely 
$\{(F,\Q)\}$ for the Streett condition
and
$\{(\emptyset,F)\}$ for the Rabin condition.

The {\em accepted language} of a nondeterministic $\omega$-automaton
$\N$ with the alphabet $\Sigma$, denoted $\L(\N)$,
 is defined as the set of infinite words $\sigma \in \Sigma^\omega$
that have at least one accepting run in $\N$.
\begin{center}
$\L(\N) \ \eqdef \ \bigl\{ \, \sigma\in \Sigma^\omega \, : \,
    \text{there exists an accepting run for $\sigma$ in $\N$} \, \bigr\}
$
\end{center}
In what follows, we write NBA to denote a nondeterministic B\"uchi automaton,
NRA for nondeterministic Rabin automata and
NSA for nondeterministic Streett automata.
Similarly, the notations DBA, DRA and DSA are used to denote deterministic
$\omega$-automata with a B\"uchi, Rabin or Streett acceptance condition.

It is well-known that the classes of languages that
can be accepted by NBA, DRA, NRA, DSA or NSA are the same.
 These languages 
are often called {\em $\omega$-regular} and represented by 
$\omega$-regular 
expressions, i.\,e., finite sums of expressions of the form
$\alpha \beta^\omega$ where 
$\alpha$ and $\beta$ are ordinary regular expressions (representing
regular languages over finite words) and the language associated with
$\beta$ is nonempty and does not contain the empty word.
In the sequel, we will identify $\omega$-regular expressions with
the induced
$\omega$-regular language.

While deterministic $\omega$-automata with Rabin and Streett acceptance
(DRA and DSA) cover the full class of $\omega$-regular languages,
DBA are less powerful
as, e.\,g., the language
$(a+b)^*a^\omega$ cannot be recognized by a DBA.
Hence,
the class of  DBA-recognizable languages
is a proper subclass of the class of $\omega$-regular languages.

Probabilistic $\omega$-automata  
can be viewed as nondeterministic 
$\omega$-automata where the transition function $\delta$
specifies probabilities for the successor states.
That is, for any state $p$ and letter $a \in \Sigma$
either $p$ does not have any $a$-successor  
or there is a probability distribution for the
$a$-successors of $p$. 

\begin{definition}[Probabilistic $\omega$-automata] 
\label{def:prob-omega-automata}
A probabilistic $\omega$-automaton is a tuple
$
   \P = (\Q, \Sigma, \delta, \mu_0, \Acc), 
$
where
 \begin{itemize}
 \item  $\Q$ is a finite nonempty set of states,
 \item  $\Sigma$ is a finite nonempty input alphabet,
 \item $\delta : \Q \times \Sigma \times \Q  \rightarrow [0,1]$
   is a \emph{transition probability function} such that 
   for all $p \in \Q$ and $a \in \Sigma$ we have:
    $\sum_{q\in Q} \delta(p,a,q) \in \{0,1\}$,
 \item $\mu_0 :\Q \to [0,1]$ is the {\em initial distribution}, i.\,e., 
   $\sum_{q\in Q} \mu_0(q)=1$,
 \item  $\Acc$ is an {\em acceptance condition}
    (as for nondeterministic $\omega$-automata).
 \end{itemize}
We refer to the states $q_0\in\Q$ where $\mu_0(q_0) >0$ as initial states.
If $p$ is a state such that $\delta(q,a,p)>0$ 
then we say that $q$ has an outgoing
$a$-transition to state~$p$.
 \end{definition}

Acceptance conditions can be defined as in the nondeterministic case.
In this paper, we just regard B\"uchi, Rabin and Streett acceptance
and use the abbreviations PBA, PRA and PSA for  
probabilistic B\"uchi automata,
probabilistic Rabin automata, and
probabilistic Streett automata, respectively.

The intuitive operational behavior of a probabilistic $\omega$-automaton
$\P$ for a given input word\linebreak
$\word = a_1 a_2 \ldots \in \Sigma^\omega$
is similar to the nondeterministic setting, except that 
all choices are resolved probabilistically:
the initial state is chosen according to the initial distribution
$\mu_0$, and if $q$ is the current state and $a$ the next input
letter then
$\P$ moves with probability
$\delta(q,a_{i+1},p)$ to state $p$.
If there is no outgoing $a$-transition from $q$, i.\,e.,
if $\sum_{p\in \Q} \delta(q,a,p)=0$,
 then $\P$ rejects. 
As in the nondeterministic case,
the resulting  
state-sequence  $\pi \ = \ 
  p_0 \,p_1 \, p_2 \ldots \in \Q^* \cup \Q^\omega$ is called
a \emph{run} for $\word$ in $\P$. 
Acceptance of a run according to a B\"uchi, Rabin or Streett
acceptance condition is defined as in the nondeterministic setting.
While acceptance of an infinite word in a nondeterministic $\omega$-automata
requires the existence of an accepting run, 
a probabilistic $\omega$-automaton accepts an infinite input word $\sigma$
if the probability for the generated sample run
to be accepting is 
``sufficiently large''.

\paragraph*{Acceptance probability and accepted language.}
Given an infinite word $\sigma \in \Sigma^\omega$, the 
\emph{acceptance probability} $\Pr^{\P}(\sigma)$ for $\sigma$ in $\P$ 
denotes the probability measure of the accepting runs
for $\sigma$ in $\P$.
The formal definition of the acceptance probability
relies on the view of an input word $\sigma \in \Sigma^\omega$  
as a {\em scheduler} when~${\cal P}$ is treated as a Markov decision process,
i.\,e., an operational model for a probabilistic system where in each state
$q$ the letters that can be consumed in $q$ are treated as actions that
are enabled in $q$. Given a word/scheduler 
$\sigma\, = \, a_1 \, a_2 \, a_3 \ldots\in \Sigma^\omega$, 
the behavior of ${\cal P}$
under $\sigma$ is given by a Markov chain ${\cal M}_{\sigma}$ 
where the states are pairs
$(q,i)$ where $q\in \Q$ stands for the current state
and $i$ is a natural number~$\geq 1$ that denotes the current 
word position. Stated differently, state $(q,i)$ in the Markov chain
$\M_{\sigma}$ stands for the configuration that $\P$ might have reached state~$q$ 
after having consumed the first $i{-}1$ letters $a_1, \ldots,a_{i-1}$
of the input word $\sigma$.
Assuming that $\delta(q,a_{i+1},\cdot)$ is  not the null function,
the transition probabilities from state $(q,i)$ are given by the distribution
$\delta(q,a_{i+1},\cdot)$, i.\,e., from state $(q,i)$ the Markov chain 
$\M_{\sigma}$ moves
with probability $\delta(q,a_{i+1},p)$ to state $(p,i+1)$.
In case that $\delta(q,a_{i+1},\cdot)=0$ then $(q,i)$ is an absorbing state,
i.\,e., a state without any outgoing transition.
The runs for $\sigma$ in ${\cal P}$ correspond 
to the paths in ${\cal M}_{\sigma}$.
We can now apply the standard concepts for Markov chains to reason about the
probabilities of infinite paths and define
the acceptance probability for the infinite word
$\sigma$
in~$\P$, denoted
 $\Pr^\P(\sigma)$ or briefly $\Pr(\sigma)$,
as the probability measure of the
accepting runs for $\sigma$ in the Markov chain $\M_\sigma$.

For the definition of the accepted language, we distinguish three
semantics for probabilistic $\omega$-automata.
The \emph{probable semantics} assigns to $\P$ the set of infinite
words $\sigma$ such that the accepting runs for $\sigma$ have positive
measure. Under the  \emph{almost-sure semantics}
a word $\sigma$ is accepted by $\P$ if almost all runs for $\sigma$
are accepting. 
(The formulation ``almost all runs have property $X$''
means that the probability measure of the runs where 
property $X$ does not hold is 0.)
The \emph{threshold semantics} relies on a fixed
threshold $\lambda$
that serves as strict lower bound for the acceptance probability
for all accepted words:
$$\begin{array}{lcl}
\L^{>0}(\P) & \eqdef & 
\bigl\{ \sigma \in \Sigma^\omega: \Pr^\P({\sigma}) > 0 \bigr\}
  \\[0.5ex]
\L^{=1}(\P) & \eqdef & 
\bigl\{ \sigma \in \Sigma^\omega: \Pr^\P({\sigma}) =1 \bigr\}
  \\[0.5ex]
\L^{>\lambda}(\P) & \eqdef & 
\bigl\{ \sigma \in \Sigma^\omega: \Pr^\P({\sigma}) > \lambda \bigr\}
  \\[0.5ex]
\end{array}
$$
Equivalence of $\omega$-automata means that their
accepted languages agree. The notion of the {\em size}, 
denoted~$|{\cal P}|$,
 of
an $\omega$-automaton ${\cal P}$ is used here as follows.
The size of a PBA is simply the
number of states. The 
size of a probabilistic Rabin or Streett 
automaton denotes the number of
states plus the number of acceptance pairs.

\begin{example}[Probabilistic B\"uchi automata (PBA)]
\label{ex:pba}
In the pictures for
PBA we attach the 
probabilities $\delta(q,a,p)$ to the $a$-labeled
edge from $q$ to $p$, provided that $0 < \delta(q,a,p) <1$.
An $a$-labeled egde from $q$ to $p$ without any probability value
indicates that $\delta(q,a,p) =1$ (in which case $p$ is the unique
$a$-successor of $q$).
Similarly, the initial distribution is depicted by attaching
the value $\mu_0(q)$ to an arrow pointing to $q$, provided that
$q$ is an initial state and $\mu_0(q)<1$.
For PBA, we depict the accepting states (i.\,e., the states
$q\in F$) by squares, non-accepting states by circles.
The PBA $\P$ over the alphabet $\Sigma = \{a,b\}$
shown in the left part of Figure \ref{fig:PBA} 
has a single initial state
$q_0$. Its B\"uchi condition is given by $F=\{q_1\}$.

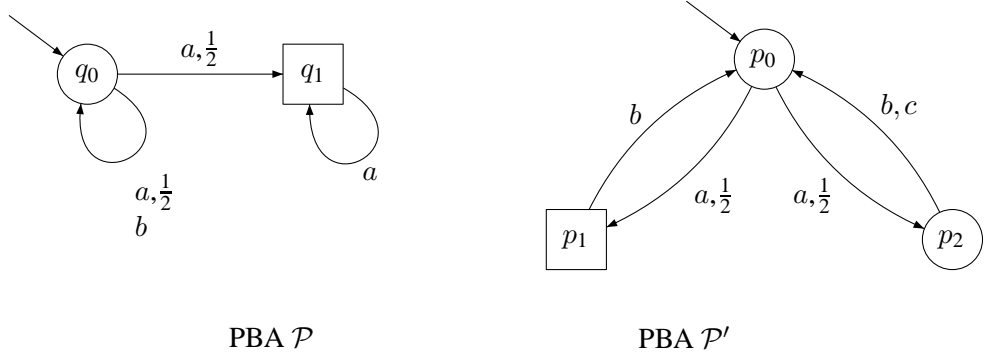
\begin{figure}[ht]
\begin{center}

\begin{picture}(130,45)(0,-45)

\node[NLangle=0.0,ilength=9.0,iangle=143.13,Nmarks=i](n0)(10.0,-10.0){$q_0$}

\node[NLangle=0.0,Nmr=0.0](n1)(40.0,-10.0){$q_1$}

\drawedge(n0,n1){{$a$,$\frac{1}{2}$}}


\drawloop[loopdiam=8.5,loopangle=-60.0](n1){{$a$}}

\drawloop[loopdiam=8.5,loopangle=-65.0](n0){
  \begin{tabular}{l}
   {$a$,$\frac{1}{2}$} \\
    $b$
  \end{tabular}}

\node[Nframe=n](text)(35.0,-45.0){PBA $\P$ }


\node[NLangle=0.0,ilength=9.0,iangle=143.13,Nmarks=i](m0)(100.0,-8.0){$p_0$}

\node[NLangle=0.0,Nmr=0.0](m1)(75.0,-32.0){$p_1$}

\node[NLangle=0.0](m2)(125.0,-32.0){$p_2$}

\drawedge[curvedepth=4.0,ELpos=50](m0,m1){{$a$,$\frac{1}{2}$}}

\drawedge[curvedepth=-4.0,ELside=r,ELpos=50](m0,m2){{$a$,$\frac{1}{2}$}}

\drawedge[curvedepth=4.0](m1,m0){$b$}

\drawedge[curvedepth=-4.0,ELside=r](m2,m0){$b,c$}

\node[Nframe=n](text)(90.0,-45.0){PBA $\P'$ }

\end{picture}

\end{center}
\caption{Examples for PBA $\P$ (left) and $\P'$ (right)}
\label{fig:PBA}
\end{figure}

Let us first observe that each word $\sigma$ that is accepted by
$\P$ must be contained in the language\linebreak $(a+b)^*a^\omega$ of the
NBA that results from $\P$ by ignoring the probabilities.
Indeed all words with only finitely many $b$'s are accepted
by $\P$ under the probable semantics, i.\,e., we have
 $\L^{>0}(\P)\, = \, (a+b)^*a^\omega$, while the almost-sure
semantics yields 
$\L^{=1}(\P) \, = \, b^*a^\omega$.
To see why, let us compute the acceptance probability for
the words $\sigma \in (a+b)^*a^\omega$.
Clearly, if $\sigma = a^\omega$ then $\Pr^{\P}(\sigma)=1$.
Suppose now that $\sigma$ contains at least one $b$
and let $k$ be the total number of
$a$'s that appear before the last $b$ in $\sigma$, i.\,e., if $\sigma = c_1 \ldots c_\ell b a^{\omega}$ then 
$k = |\{ i\in \{1,\ldots,\ell\}: c_i = a\}|$.
With probability $2^{-k}$, the current state is
$q_0$ after reading $c_1 \ldots c_\ell$.
But then $\P$ can read $b$ and will almost surely move to $q_1$
when reading the suffix $a^{\omega}$.
Thus, $\Pr^{\P}(\sigma) = 2^{-k}$ which yields that
$\sigma \in \L^{=1}(\P)$ iff $k=0$ (i.\,e., if $\sigma \in b^*a^{\omega}$)
and that
all words in $(a+b)^*a\omega$ belong to $\L^{>0}(\P)$.

Regard
 the PBA $\P'$ over the alphabet $\Sigma = \{a,b,c\}$
shown in the right part
of Figure \ref{fig:PBA}. 
Let us first observe 
that the underlying nondeterministic B\"uchi automaton 
(NBA) 
that we obtain by ignoring the probabilities has an accepting run
for each infinite word in $(ab+ac)^\omega$ with infinitely many
$b$'s, no matter whether there are only finitely $c$'s or infinitely many
$c$'s. Thus, the accepted language of the NBA
is $\bigl( (ac)^*ab\bigr)^\omega$. 
This language is different from the accepted language
of the PBA $\P'$ under the probable and almost-sure
 semantics:
$$\L^{>0}(\P')  =  
   (ab + ac)^*(ab)^{\omega}, \ \ \ 
  \L^{=1}(\P')  =  (ab)^{\omega}.
$$
Clearly, all accepted words $\sigma \in \L^{>0}(\P')$ 
belong to $((ac)^*ab)^\omega$.
Any word $\word$ in $(ab + ac)^{\omega}$
with infinitely many $c$'s is rejected by $\P'$  
as  almost all runs for $\word$ are finite and end in state $p_1$,
where the next input symbol 
is $c$ and cannot be consumed in state $p_1$.
Thus, 
$\L^{>0}(\P') \ \subseteq \  (ab + ac)^*(ab)^{\omega}$.
Given an input word $\sigma \in (ab + ac)^*(ab)^{\omega}$,
say $\sigma = x(ab)^\omega$ where $x\in (ab+ac)^*$, then
with positive probability $\P'$ generates the run fragment
$p_0\, p_2 \, p_0 \, p_2 \ldots p_0 \, p_2 \, p_0$ when
reading $x$. For the remaining suffix $(ab)^\omega$, $\P'$ can always
consume the next letter and almost surely $\P'$
will visit $p_1$ and $p_2$ infinitely
often. This yields $\Pr^{\P'}(\sigma) >0$ and $\sigma \in \L^{>0}(\P')$.

Clearly, we have $\L^{=1}(\P')\subseteq \L^{>0}(\P')$.
Using an argument as above, it is clear that 
no word in $\L^{=1}(\P')$ contains letter $c$.
The runs for the word $(ab)^{\omega}$ will almost surely
visit state $p_1$ infinitely often.
This yields $\L^{=1}(\P') = (ab)^{\omega}$.

The precise acceptance probability for 
$\sigma \in \{a,b,c\}^\omega$
is as follows. If \hbox{$\sigma \notin (ab+ac)^*(ab)^\omega$} then
$\Pr^{\P'}(\sigma) = 0$. If $\sigma \in (ab+ac)^*(ab)^\omega$ 
and letter $c$ appears $k$ times in $\sigma$ then
$\Pr^{\P'}(\sigma) = 2^{-k}$.
Thus, e.\,g., for threshold $0.1$, the accepted language
$\L^{>0.1}(\P')$ consists of all words
$\sigma \in (ab+ac)^*(ab)^\omega$ that contain three or fewer
$c$'s.
\hfill$\diamond$
\end{example}

\section{Expressiveness and efficiency of PBA}
\label{section:express}

In the following three sections, we put the focus on
probabilistic $\omega$-automata with the probable semantics.
Results for the almost-sure and threshold semantics are summarized
in Section \ref{section:almost}.
Unless stated differently, we simply say PBA to denote a PBA with
the probable semantics.

We start with a discussion on the expressiveness and efficiency
of PBA compared
to their nondeterministic counterparts.
At the end of this section, we will show that 
as in the nondeterministic case, B\"uchi acceptance
is as powerful as Streett and Rabin acceptance.

\paragraph*{PBA and $\omega$-regular languages.}
DBA can be viewed as special instances of PBA 
(we just have to assign probability 1 to all edges
in the DBA and deal with the initial distribution that
assigns probability 1 to the unique initial state).
As the language  $(a+b)^*a^\omega$
is recognizable by a PBA with the probable semantics
(see Example \ref{ex:pba}), PBA are strictly more expressive
than DBA, i.\,e.,
the class of DBA-recognizable languages is a proper
subclass of the class of languages $\L^{>0}(\P)$ for some PBA $\P$.
Indeed all $\omega$-regular languages can be
represented by a PBA with the probable semantics:

\begin{lemma}[From NBA to PBA under the probable semantics]%
\label{lemma:NBA-2-PBA}%
\hspace*{-3pt}For each NBA $\N$ 
there exists a PBA ${\cal P}$ 
such that $\L^{>0}(\P)=\L(\N)$.
\end{lemma}

\begin{proof}
A  transformation from NBA into an equivalent PBA
is obtained by using NBA that are {\em de\-ter\-mi\-nis\-tic-in-limit}.
These are NBA such that 
$\delta(q,a,p)\in \{0,1\}$
for all states $p$ and $q$ that are reachable from some accepting state
and all letters $a\in \Sigma$.
That is, as soon as an accepting state has been reached the
behavior from then on is deterministic.
Courcoubetis and Yannakakis
\cite{Courcoubetis:1995:CPV} presented some kind of powerset
construction which turns a given NBA ${\cal N}$
into an equivalent NBA $\N_{\text{det}}$ that is
deterministic-in-limit.
If we now resolve the 
nondeterministic choices
in~$\N_{\text{det}}$ by uniform distributions\footnote{If $q$ is a state
in $\N_{\text{det}}$ and $a\in \Sigma$ such that
$q$ has $k$ $a$-successors $q_1,\ldots,q_k$
then we define $\delta(q,a,q_i) = \frac{1}{k}$ for $1 \leq i \leq k$ and
$\delta(q,a,p)=0$ for all states $p\notin \{q_1,\ldots,q_k\}$. Similarly,
if $\Q_0$ is the set of initial states in $\N_{\text{det}}$ and 
$\Q_0$ is nonempty then
we deal with the initial distribution $\mu_0$ that assigns probability
$1/|\Q_0|$ to each state in $\Q_0$.}
then $\N_{\text{det}}$ becomes a PBA that accepts the same language
as $\N$ (and $\N_{\text{det}}$). 
\end{proof}

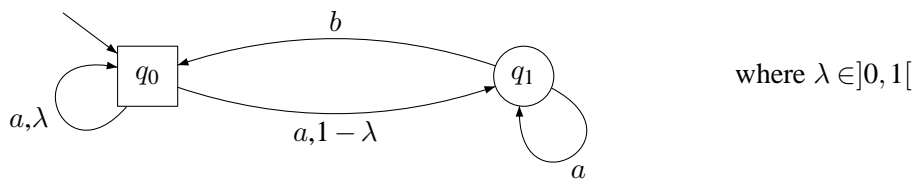
\begin{figure}[ht]
\begin{center}

\begin{picture}(130,25)(0,-25)

\node[NLangle=0.0,ilength=9.0,iangle=143.13,Nmarks=i,Nmr=0.0](n0)(20.0,-10.0){$q_0$}

\node[NLangle=0.0](n1)(70.0,-10.0){$q_1$}

\drawedge[curvedepth=-5.0,ELside=r](n0,n1){$a$,$1-\lambda$}

\drawedge[curvedepth=-5.0,ELside=r](n1,n0){$b$}

\drawloop[loopdiam=8.5,loopangle=200.0](n0){$a$,$\lambda$}

\drawloop[loopdiam=8.5,loopangle=-60.0](n1){$a$}

\node[Nframe=n](lambda)(110.0,-10.0){where $\lambda \in ]0,1[$}

\end{picture}

\end{center}
\caption{PBA $\P_\lambda$
   accepts a non-$\omega$-regular language}
\label{fig:PBA-not-reg}
\end{figure}

We now address the question whether
each PBA can be transformed into an equivalent
NBA. Surprisingly, this is not the case,
as there are PBA where the accepted language
is not $\omega$-regular. 
An example for a PBA $\P = \P_\lambda$
where the acepted language under the
probable semantics is not $\omega$-regular
is given in
Figure~\ref{fig:PBA-not-reg}.

Here, $\lambda$ is an arbitrary
real number in the open interval $]0,1[$.

\begin{lemma}
\label{lemma:P-lambda-not-reg}
The language of the 
 PBA $\P_\lambda$ under the probable semantics is not NBA-recognizable,
i.\,e.,
$\L^{>0}(\P)$ is not $\omega$-regular.
\end{lemma}

\begin{proof}
The PBA $\P_\lambda$ accepts the language
$$\L^{>0}(\P_\lambda) \ = \ \Bigl\{ \, a^{k_1}b a^{k_2}b a^{k_3}b \ldots :
    \prod_{i=1}^\infty \bigl( 1- \lambda^{k_i} \bigr) \, > \, 0  \ \Bigr\}.
$$
The convergence condition 
which requires the infinite product over the values
$1- \lambda^{k_i}$ to be positive
can easily be shown to be non-$\omega$-regular, i.\,e.,  $\L^{>0}(\P_\lambda)$ 
cannot be recognized by an NBA.

To see that, indeed, $\L^{>0}(\P_{\lambda})$ agrees with the above language, 
let us first observe that all words
in $\L^{>0}(\P_\lambda)$ must contain infinitely many $b$'s.
Note that if a input word $\sigma$ ends with the suffix $a^{\omega}$
then almost all infinite runs for $\sigma$ will eventually enter
state $q_1$ and stay there forever.
As $\P_\lambda$ cannot consume two consecutive $b$'s,
all words in $\L^{>0}(\P_\lambda)$  have the form
$a^{k_1}b a^{k_2}b a^{k_3}b \ldots $ where
$k_1,k_2,\ldots$ is a sequence of positive natural numbers.
We now show that
$$\Pr^{\P_\lambda}(a^{k_1}b a^{k_2}b a^{k_3}b \ldots ) \ = \
 \prod_{i=1}^\infty \bigl( 1- \lambda^{k_i} \bigr). 
$$
The factors $1-\lambda^{k_i}$ stand for the probability
to move from state $q_0$ to $q_1$ when reading the subword $a^{k_i}$.
With the remaining probability $\lambda^{k_i}$, 
the automaton $\P_\lambda$
stays in state $q_0$, but then letter $b$ at position
$k_1 + \dots + k_i + i$ of the input word
$a^{k_1}b a^{k_2}b a^{k_3}b \ldots$
cannot be consumed and
$\P_{\lambda}$ rejects.
Hence, the probability for run fragments
of the form $q_0 \ldots q_0 q_1 \ldots q_1 \, q_0$ 
that are generated while reading the subword $a^{k_i}b$ 
is precisely $1-\lambda^{k_i}$.
This yields that the infinite product of these values
agrees with the acceptance probability for 
 the input word $a^{k_1}b a^{k_2}b a^{k_3}b \ldots$.%
\end{proof}

As a consequence of  
Lemma \ref{lemma:NBA-2-PBA} and 
Lemma \ref{lemma:P-lambda-not-reg}
we get that PBA with the probable semantics are more powerful
than NBA. This result should be contrasted to the
case of finite automaton where the probable semantics turns 
PFA into ordinary NFA, and thus, PFA with the probable semantics
represent exactly the class of $\omega$-regular languages.

\begin{corollary}
The class of languages that are accepted by a PBA 
strictly subsumes the class of $\omega$-regular languages.
\end{corollary}

The PBA $\P_{\lambda}$
can also serve to illustrate that the probable semantics
is sensititive to modifications of the transition probabilties.
Consider two values $\lambda$ and $\nu \in ]0,1[$ with
$\lambda < \nu$. For any sequence
$(k_i)_{i \geq 1}$ of natural numbers $k_i$ 
where 
the infinite product over
the values $1-\nu^{k_i}$ converges to some positive value,
also the infinite product over
the values $1-\lambda^{k_i}$ is positive, as we have
$1-\nu^{k_i} < 1-\lambda^{k_i}$.
Thus,
$\L^{>0}(\P_\nu) \subseteq \L^{>0}(\P_\lambda)$.
However, whenever $\lambda < \nu$ then
$\L^{>0}(\P_\nu)$ 
is a proper sublanguage
$\L^{>0}(\P_\lambda)$ as
there are sequences
$(k_i)_{i \geq 1}$ such that
the product of the values $1-\lambda^{k_i}$ converges to some positive
real number, 
while the product of the values $1-\nu^{k_i}$ has value 0
\cite{BaiBerGroe08}.
Hence:

\begin{lemma}
If $\lambda < \nu$ then
$\L^{>0}(\P_\nu) \, \not= \, \L^{>0}(\P_\lambda)$.
\end{lemma}
 
Thus, the languages of PBA are sensitive to the distributions
for the successor states. That is, if we are given a PBA and
modify the nonzero transition probabilities then also the accepted language
might change.
This property is surprising since the definition of the
accepted language just relies on a qualitative criterion: the
acceptance probability must be positive, but might be arbitrarily small.
This should be opposed to the verification of finite-state Markov decision
processes where it is known that whether or not a given linear time property
holds with positive probability just depends on the underlying graph,
but not on the concrete transition probabilities. 

\paragraph*{\bf Efficiency.}
The transformation from NBA into an equivalent PBA with the probable
semantics described in the 
proof of Lemma \ref{lemma:NBA-2-PBA}
relies on a certain
powerset construction that
turns a given NBA $\N$ into an NBA that is deterministic-in-limit and
can be interpreted as a PBA. This transformation 
 can cause an exponential blow-up.
In fact, in the worst-case, the  exponential blow-up cannot be avoided
for the transformation from NBA to PBA as
there are families $(L_n)_{n \geq 1}$ of
$\omega$-regular languages
that are accepted by NBA of linear size, while each PBA for
$L_n$ has $\Omega(2^n)$ states.
An example for such a family of languages 
is 
$$L_n \ = \ ((a+b)^*a (a+b)^n c)^\omega.
$$
Language $L_n$ is recognizable by an NBA with $n{+}1$ states
which guesses nondeterministically for any word position $i$
where the input word contains an $a$ 
whether letter $c$ will appear at word position $i{+}n$.
Since there is no upper bound on the distance between
the word positions of the $c$'s in the words in $L_n$, any
PBA for $L_n$ needs to store the positions of letter 
$a$ among the last $n$ letters (see \cite{BG-lics05}). 
Hence, the 
size of any PBA for $L_n$ is exponential.
Vice versa, there are also examples for $\omega$-regular 
languages where probabilism allows for a more compact representation
than nondeterminism. Let
$$L_n' \ \eqdef \ \bigl\{ \, xy^\omega\, : \,
   x,y\in \{a,b\}^*, |y|=n \, \bigr\}.
$$

\begin{lemma}
Each NSA for $L_n'$ has
$2^n/n$ or more states
in each NSA for $L_n'$, while 
there exist PBA $\P_n$ consisting of ${\cal O}(n)$ states with
$\L^{>0}(\P_n) = L_n'$.
\end{lemma}

\begin{proof}
The lower bound $2^n/n$ for the number of states in any NSA
for $L_n'$ is obtained by verifying that given two
words $y_ = c_1 c_2 \ldots c_n$ and $z = d_1 d_2 \ldots d_n$
of length $n$ such that 
\[d_1 d_2 \ldots d_n \notin \bigl\{ \, c_i c_{i{+}1} \ldots c_n c_1 \ldots
   c_{i{-}1} : 1 \leq i \leq n \, \bigr\}\]
then the ``accepting cycles'' for the words $y^{\omega}$, $z^{\omega}\in L_n'$
do not intersect. 

It remains to show the existence of PBA of linear size 
for $L_n'$.
Let $\P_n$ be the following PBA.
The states of $\P_n$ are
$1_a,\ldots,n_a,1_b,\ldots,n_b$.
Thus, $\P_n$ has~$2n$ states.
States $1_a$ and $1_b$ are initial, both have
probability 0.5 under the initial distribution. 
All states are accepting. (Thus, any infinite run in $\P_n$ is accepting.)
$\P_n$ has the following transitions.
From any state $k_a$ with $1 \leq k < n$ 
there is an $a$-transition to state $(k{+}1)_a$ 
 and a $b$-transition to state $(k{+}1)_b$. All these transitions 
have probability 0.5.
All states, except for state $n_b$, have an $a$-transition to state~$1_a$.
These transitions have 
probability 0.5, except for the transition
from~$n_a$ to~$1_a$ which has probability 1.
Similarly,
from any state $k_b$ with $1 \leq k < n$ 
there is an $a$-transition and a $b$-transition to state $(k{+}1)_b$
with probability 0.5.
All states, except for state $n_a$, have a $b$-transition to state $1_b$ with probability 0.5 except for 
state $n_b$ which has a $b$-transition to $1_b$ with probability~1. 

The idea of this construction is as follows.
While scanning an infinite input word 
$$\sigma = c_1 c_2 c_3 \ldots
\in \{a,b\}^\omega,$$
$\P_n$ chooses at random word positions $i$ by moving to state $1_a$
(if $c_i=a$) or state $1_b$ (if $c_i=b$)
and checks whether $c_{n{+}i} = c_i$ 
via following 
the path 
\[1_{c_i} \overto{c_{i{+}1}}   2_{c_i} 
          \overto{c_{i{+}2}}
               \cdots 
          \overto{c_{i{+}n{-}1}} n_{c_i}
\]
and rejecting (if $c_{n{+}i} \not= c_i$)
or returning to state $1_{c_i}$ (if $c_{n{+}i} = c_i$)
and choosing the next word position~$j$, and so on.
If $\sigma \notin L_n'$ then there are infinitely many
word positions $i$ such that $c_{n{+}i}\not= c_i$ and
almost surely $\P_n$ will pick such a word position and
reject in state $n_{c_i}$.
If $\sigma = c_1 c_2 c_3 \ldots \in L_n'$ then 
there exists some index $\ell$ such that
$c_i = c_{n{+}i}$ for all $i \geq \ell$.
After reading the $\ell$-th letter,
$\P_n$ will be in state $1_{c_\ell}$ with
probability $\geq 2^{-\ell}$.
From then on, $\P_n$ will never reject and the resulting
runs are accepting. Hence, $\Pr^{\P_n}(\sigma) > 0$.
\end{proof}

\paragraph*{\bf Streett and Rabin acceptance.}
The three types of
probabilistic $\omega$-automata (B\"uchi, Rabin, Streett)
are equally expressive.
As the B\"uchi acceptance condition can be
rewritten as a Rabin or Streett acceptance condition, 
each PBA can be viewed as a PRA 
or as a PSA with the same accepted language.
But we can establish a stronger result stating that each PBA can
be transformed into a 0/1-PRA which means a PRA $\P_R$ 
such that for each word $\sigma$, the acceptance probability
for $\sigma$ is either 0 or 1.
This result can be viewed
as the probabilistic analogue to the well-known fact that
each NBA can be transformed into an equivalent
deterministic Rabin automaton. 
The idea for this transformation 
is to design
a 0/1-PRA $\P_R$ 
that generates up to $n$ sample runs of $\P$ and checks whether
at least one of them is accepting, where $n$ is the number of states
in $\P$. If so then $\P_R$ accepts, otherwise it rejects.
For the details of this construction 
we refer to \cite{BaiBerGroe08,GroesserThesis}.

\begin{theorem}[From PBA to 0/1-PRA]
\label{theorem:0-1-PRA}
For each PBA $\P$ there exists a
0/1-PRA $\P_R$ such that
$$\L^{>0}(\P) = \L^{>0}(\P_R).$$
\end{theorem}

Vice versa, there are polynomial transformations from
PRA and PSA to PBA:

\begin{theorem}[Polynomial transformations from 
  PBA\;to PRA\;and PSA]\ \\[-3mm]
\label{theorem:equiv-PBA-PSA-PRA} 
\begin{enumerate}
\item [{\rm (a)}]
Given a PRA ${\cal P}_R$ with 
$\ell$ acceptance pairs  there exists
a PBA $\P$ 
of size ${\cal O}(\ell|{\cal P}_R|)$
such that\\
$\L^{>0}(\P) =   \L^{>0}(\P_R)$.
\item [{\rm (b)}]
Given a PSA ${\cal P}_S$ with 
$\ell$ acceptance pairs  there exists
a PBA of size
${\cal O}(\ell^2|{\cal P}_S|)$
such that\\ $\L^{>0}(\P) = \L^{>0}(\P_S)$.
\end{enumerate}
\end{theorem}

The transformation from PRA to PBA is roughly the same
as in the nondeterministic case.
The construction of a PBA of size 
${\cal O}(\ell^2|{\cal P}_S|)$
from a given PSA~$\P_S$, however,
crucially relies on the probabilistic semantics.
In fact, it is worth noting that in the nonprobabilistic case
it is known (see \cite{SafraVardi89})
that there are families $(L_n)_{n\geq 0}$ of
languages $L_n \subseteq \Sigma^\omega$ that are recognizable
by nondeterministic Streett automata of size
${\cal O}(n)$, while each 
nondeterministic B\"uchi automaton for~$L_n$ 
has~$2^n$ or more states. Thus, the
polynomial transformation from Streett to B\"uchi acceptance
is specific for the probabilistic case.

\section{Composition operators for PBA}
\label{section:comp}

The most important composition operators for any class of languages over
infinite words are the standard set operations union, intersection
and complementation.
In fact, the class of PBA-recognizable languages is closed
under all three operations.

\begin{theorem}
The class of languages $\L^{>0}(\P)$ for some PBA $\P$
is closed under union, intersection and complementation.
\end{theorem}

Given two PBA $\P_1$ and $\P_2$ over the same alphabet
with initial distributions~$\mu_1$ and~$\mu_2$, respectively, then a PBA $\P$ 
for the
language $\L^{>0}(\P_1)\cup \L^{>0}(\P_2)$
can be obtained by
the disjoint union of $\P_1$ and $\P_2$ with the initial distribution
$\mu(q) = \frac{1}{2}\mu_i(q)$ if $q$ is a state in $\P_i$.
If $F_1$ and $F_2$ are the sets of accepting states in $\P_1$ and $\P_2$,
respectively, then $\P$ requires to visit $F_1 \cup F_2$ infinitely often.

An operator for PBA with the probable semantics
that realizes {\em intersection} can be designed by
reusing ideas that are known 
for NBA. Given two PBA~$\P_1$ and~$\P_2$ over the same
alphabet, we use a product construction
$\P_1 \times \P_2$ (which runs~$\P_1$ and~$\P_2$ in parallel)
and equip $\P_1\times \P_2$ with a Streett acceptance
condition consisting of two
acceptance pairs. One of the acceptance pairs 
requires that an accepting state of $\P_1$ is visited
infinitely often, the other one stands for the acceptance
condition of $\P_2$.
This PSA $\P_1 \times \P_2$ can then be transformed into an equivalent PBA
(part (b) of Theorem \ref{theorem:equiv-PBA-PSA-PRA}).

The most interesting operator is {\em complementation}.
Given a PBA $\P$ with 
$L = \L^{>0}(\P) \subseteq \Sigma^\omega$, the idea for
the construction of a PBA $\overline{\P}$ for the language
$\overline{L} = \Sigma^\omega \setminus L$ is somehow similar to the
complementation of NBA via Safra's determinisation operator \cite{Safra88}
and relies on the 
transformations sketched in Figure~\ref{fig:complementation}.
\begin{figure}[ht]
\begin{center}
\begin{tabular}{ccccccc}
\begin{tabular}{c}
  PBA $\P$ with \\[-0.2ex]
  $L = \L^{>0}(\P)$
\end{tabular}
& \ \ {\Large $\leadsto$} \ \ & 
\begin{tabular}{c}
0/1-PRA $\P_R$ \\[-0.2ex]
 for $L$ 
\end{tabular} 
& \ \ {\Large $\leadsto$}\ \ & 
\begin{tabular}{c}
0/1-PSA $\P_S$ \\[-0.2ex]
  for $\overline{L}$ 
\end{tabular}
& \ \ {\Large $\leadsto$} \ \ & 
\begin{tabular}{c}
PBA $\overline{\P}$ with \\[-0.2ex]
   $\L^{>0}(\overline{\P}) = \overline{L}$ 
\end{tabular}
\end{tabular}
\end{center}
\caption{Complementation of a PBA}
\label{fig:complementation}
\end{figure}
In the first step we apply the transformation
mentioned in Theorem \ref{theorem:0-1-PRA},
while the last step relies on
part (b) of Theorem \ref{theorem:equiv-PBA-PSA-PRA}. 
Recall that
 a 0/1-PRA denotes a PRA~$\P_R$ 
where the acceptance probabilities for all words
are 0 or 1, i.\,e.,
 \hbox{$\Pr^{\P_R}(\sigma) \in \{0,1\}$}
 for each word
$\sigma \in \Sigma^\omega$.
Thus, $\L^{>0}(\P_R) = \L^{=1}(\P_R)$ and 
for transforming the 0/1-PRA $\P_R$ 
into a 0/1-PSA $\P_S$ for the complement
of $\L^{>0}(\P_R)$
we may simply use the duality
of Rabin and Streett acceptance.
That is, syntactically~$\P_R$ and~$\P_S$ agree
 (but $\P_S$ is viewed as a Streett and $\P_R$ as a Rabin automaton). 
The size of the resulting PBA $\overline{\P}$
for $\overline{L}$ can be exponentially
larger than the size of $\P$ due to the powerset construction used 
in the generation of a 0/1-PRA.

\section{Decision problems for PBA}
\label{section:dec}

For many applications of automata-like models,
it is important to have (efficient) decision algorithms
for some fundamental problems, like checking emptiness
or language inclusion.
For instance, the automata-based approach \cite{VardiWolper86}
for verifying $\omega$-regular properties of a nondeterministic
finite-state system relies on a reduction to the 
emptiness problem for NBA.
Unfortunately, the emptiness problem and various other classical
decision problems for automata cannot be solved algorithmically
for PBA:

\begin{theorem}[Undecidability of PBA]
\label{theorem:undec}
The following problems are undecidable:
\begin{itemize}
\item {\bf \em emptiness:}
  given a PBA $\P$, does $\L^{>0}(\P) =\emptyset$ hold?
\item {\bf \em universality:}
  given a PBA $\P$ with the alphabet $\Sigma$, 
  does $\L^{>0}(\P) =\Sigma^\omega$ hold?
\item {\bf \em equivalence:}
  given two PBA $\P_1$ and $\P_2$, 
  does $\L^{>0}(\P_1) =\L^{>0}(\P_2)$ hold?
\end{itemize}
\end{theorem}

To prove undecidability of the emptiness problem,
we provided in \cite{BaiBerGroe08} a reduction from 
a variant of the
emptiness problem for probabilistic finite automata 
(PFA) which has been shown to be undecidable~\cite{condon-AI03}.
Undecidability of the universality problem then follows
by the effectiveness of complementation for PBA.
Undecidability of the PBA-equivalence problem is an immediate
consequence of the undecidability of the emptiness problem
(just consider $\P_1=\P$ and $\P_2$ a PBA for the empty language).

A consequence of Theorem \ref{theorem:undec} is that
PBA are not appropriate for verification algorithms.
Consider, e.\,g., a finite-state transition system $\T$
and suppose that a linear-time property $p$ to be verified
for $\T$ is specified by a PBA $\P$ in the sense 
that~$\L^{>0}(\P)$ represents all infinite behaviors where property $p$ holds.
(Typically $p$ is a language over some alphabet $\Sigma = 2^{\text{AP}}$ where
$\text{AP}$ is a set of atomic propositions and the states in
$\T$ are labeled with subsets of $\text{AP}$.)
Then, the question whether 
all traces of $\T$ have property $p$ is reducible to the
universality problem for PBA and therefore undecidable.
Similarly, the question whether $\T$ has at least one trace
where $p$ holds is reducible to the emptiness problem for PBA and
therefore undecidable too.

Another important consequence of Theorem \ref{theorem:undec}
is that it yields the undecidability of the verification
problem for partially observable Markov decision processes (POMDPs)
against $\omega$-regular properties. 
POMDPs provide an operational model for stochastic 
systems with non-observable behaviors. 
They play a central role in many application areas
such as mobile robot navigation,  probabilistic planning task, 
elevator control, and so on. See, e.\,g.,
 \cite{Sondik71thesis,Monahan82,PapTsi87,Lovejoy91}.
The syntax of a POMDP can be defined as for probabilistic
$\omega$-automata, except that 
the acceptance condition has to be replaced with an equivalence
relation $\sim$ on the states which formalizes which states 
cannot be distinguished from outside.
The elements in the alphabet $\Sigma$ are viewed as action names.
The goal is then to design a {\em scheduler} $\S$
that chooses the actions for
the current state and ensures that a certain condition $X$ holds 
when the choices between different enabled actions in the 
POMDP $\M$ are resolved by $\S$.
For his choice the scheduler may  
use the sequence of equivalence classes that have been passed
to reach the equivalence class of the current state.
That is, the
scheduler is supposed to observe the equivalence classes, but not the
specific states.
(Such schedulers are sometimes called ``partial-information schedulers''
or ``observation-based schedulers''.)

The emptiness problem for PBA is a special instance for
the scheduler-syn\-the\-sis problem for\linebreak POMDPs.
Given a PBA $\P = (\Q,\Sigma,\delta,\mu_0,F)$, 
we regard the POMDP $\M = (\Q,\Sigma,\delta,\mu_0,\sim)$ where
$\sim$ identifies all states and ask for the existence of 
a scheduler that
ensures that $F$ will be visited infinitely often 
with positive probability.
We first observe that the infinite words over $\Sigma$ can be
viewed as schedulers for $\M$, and vice versa.
Hence, $\L^{>0}(\P)$ is nonempty if and only if there is a scheduler
$\S$ such that $\Pr^{\M}_{\S}(\Box \Diamond F) > 0$, where
$\Pr^{\M}_{\S}(\Box \Diamond F)$ denotes the probability that $\M$ 
visits $F$ infinitely often when $\S$ is used to schedule the actions
in $\M$.
Similarly, the universality problem for PBA can be viewed as a special
instance of the problem where we are given a POMDP $\M$
 and a set $F$ of states and ask 
for the existence of a scheduler $\S$ such that
$\Pr^{\M}_{\S}(\Diamond \Box F)=1$
where $\Pr^{\M}_{\S}(\Diamond \Box F)$ denotes the probability 
that $\M$ under scheduler $\S$
eventually enters $F$ and never leaves $F$ from this moment on.
Thus: 

\begin{theorem}[Undecidability results for POMDPs]
\label{theorem;POMDP-undec}\ \\
The following problems are undecidable:
\begin{itemize}
\item given a POMDP $\M$ and a set $F$ of states, 
   decide whether $\exists \S. \, \Pr^{\M}_{\S}(\Box \Diamond F) > 0,$
\item given a POMDP $\M$ and a set $F$ of states, 
   decide whether $\exists \S. \, \Pr^{\M}_{\S}(\Diamond \Box F) =1.$
\end{itemize}
\end{theorem}

The result of Theorem \ref{theorem;POMDP-undec} is remarkable
since the corresponding questions for fully observable Markov decision
processes (i.\,e., POMDPs where the $\sim$-equivalence classes are singletons)
are decidable in polynomial time.

\section{The almost-sure and threshold semantics}
\label{section:almost}

So far, we concentrated on the probable semantics of 
probabilistic $\omega$-automata. We will briefly summarize
the main results on the almost-sure and threshold semantics.

PBA with the almost-sure semantics are less expressive
than PBA with the probable semantics.
They even do not cover the full class of $\omega$-regular languages.
For instance, the $\omega$-regular language $(a+b)^*a^\omega$ cannot be
recognized by a PBA with the almost-sure semantics. 
Since the complement $(a^*b)^\omega$ of this language is 
recognizable by a deterministic B\"uchi automaton (and therefore also by a
PBA with the almost-sure semantics),
PBA with the almost-sure semantics are not closed under
complementation.
Furthermore, there are PBA where the almost-sure semantics 
yields a non-$\omega$-regular language. An example is the 
language 
$$L \ = \ \Bigl\{ \, a^{k_1}b a^{k_2}b a^{k_3}b \ldots :
    \prod_{i=1}^\infty \bigl( 1- \lambda^{k_i} \bigr) \, = \, 0  \ \Bigr\}
$$
which can be shown to be
 recognizable by a PBA with
the almost-sure semantics. 
However, the class of languages $\L^{=1}(\P)$ for some PBA $\P$
is closed under union and intersection. 
For PBA with the almost-sure semantics, the emptiness and universality
problem are decidable. 
Indeed one can even show that given a POMDP $\M$
and a set $F$ of states in $\M$ then the questions 
\begin{center}
\begin{tabular}{l}
does there exists
a scheduler $\S$ such that $\Pr^{\M}_{\S}(\Box \Diamond F) = 1$?\\[0.5ex]
does there exists
a scheduler $\S$ such that $\Pr^{\M}_{\S}(\Diamond \Box F) >0$?\\[0.5ex]
\end{tabular}
\end{center}
are decidable by a certain powerset construction. 
Using the above mentioned fact that PBA can be viewed
as special instances of POMDPs, one obtains the decidability of the emptiness
and universality problem for PBA with the almost-sure semantics.

It should be noticed that the above results
on the almost-sure semantics are specific for the B\"uchi
acceptance condition.
For Rabin or Streett acceptance,
the almost-sure semantics is as expressive as
the probable semantics.
This is a consequence of
 Theorems \ref{theorem:0-1-PRA}
and  \ref{theorem:equiv-PBA-PSA-PRA} which show
that
PRA with the almost-sure semantics are as
expressive as  PRA (and PBA) with the probable semantics.
Thus, the emptiness,
universality and equivalence problems 
for PRA with the almost-sure semantics 
are undecidable.

The threshold semantics is more powerful than the
probable semantics. Indeed for each PBA $\P$ and threshold $\lambda$
there exists a PBA $\P'$ such that $\L^{>0}(\P)=\L^{>\lambda}(\P')$.
Furthermore, there are transformations to stretch and
relax acceptance probabilities which yields that whenever
$\lambda,\nu \in ]0,1[$ and $\P$ is a PBA then there exists
a PBA $\P'$ such that $\L^{>\lambda}(\P)=\L^{>\nu}(\P')$.
That is, all thresholds define the same class of
languages. Using known results on the expressiveness of probabilistic
finite automata (PFA) \cite{Rabin63,paz71}, 
one can show that there are threshold 
languages  $\L^{>\lambda}(\P)$ that cannot be recognized by
PBA with the probable semantics.
The undecidability of all relevant algorithmic problems for PBA with
the threshold semantics is clear from the undecidability of 
corresponding problems for PFA \cite{condon-AI03}.
As far as we know, closure properties under
composition operators have not yet been studied for
PBA with the threshold semantics.

\section{Conclusion}
\label{section:conc}

We gave a summary of the fundamental properties of 
probabilistic acceptors for infinite words 
formalized by probabilistic $\omega$-automata
with B\"uchi, Rabin or Streett acceptance conditions.
The results show some major differences to  
nondeterministic
(or alternating) $\omega$-automata 
concerning the expressiveness,
efficiency and decidability.

Beside being of theoretical interest, we believe that PBA 
could be useful in several application areas.
We briefly sketched the connection between probabilistic
$\omega$-automata and POMDPs.
Since PBA arise as special instance of POMDPs all negative
results for PBA (undecidability) carry over from PBA to POMDP.
Vice versa, it seems that for many 
algorithmic problems for POMDPs, algorithmic solutions for 
probabilistic $\omega$-automata (e.\,g., PBA with the almost-sure semantics)
can be combined with
standard algorithms for (fully observable) Markov decision processes
to obtain an algorithm that solves the analogous problem for POMDPs. 
Another application of probabilistic $\omega$-automata
 is run-time verification
where special types of PBA can serve as probabilistic monitors
\cite{CSistlaV08}.
Given the wide range of application areas of probabilistic finite automata,
there might be various other applications of probabilistic
$\omega$-automata.
For instance, the concept of probabilistic $\omega$-automata
is also related to partial-information 
games with $\omega$-regular winning 
objectives \cite{ChatDoyHenzRask06} or
could serve as starting point for studying quantum automata over
infinite inputs, in the same way as PFA yield the basis
for the definition of quantum finite automata
\cite{KondacsWatrous97,AmbainisFreivalds98}.

\bibliographystyle{eptcs} 
\bibliography{biblio_thesis}

\end{document}